\documentstyle[aps,manuscript]{revtex}
\begin{document}

\title{Signatures of Topological Defects}

\author{Veniamin Berezinsky}
\address{INFN, Laboratori Nazionali del Gran Sasso, I-67010 Assergi (AQ), Italy
\\ and Institute for Nuclear Research , Moscow, Russia}
\author{Pasquale Blasi}
\address{Department of Astronomy \& Astrophysics and\\
E. Fermi Institute, The University of Chicago,\\
5640 S. Ellis Av., Chicago, IL 60637, USA}
\author{Alexander Vilenkin}
\address{Institute of Cosmology, Department of Physics and Astronomy,\\
Tufts University, Medford, MA 02155, USA}

\maketitle

\begin{abstract}

We argue that due to various restrictions cosmic strings and
monopole-string networks are not
likely to produce the observed flux of ultra-high energy cosmic rays
(UHECR).  Among the 
topological defects studied so far, the most promising UHECR sources
are necklaces and monopolonia. Other viable sources which are 
similar to topological defects are relic superheavy particles. All these 
sources have an excess of pions (and thus photons) over nucleons at 
production. We demonstrate that in the case of necklaces the diffuse proton 
flux can be larger than photon flux, due to absorption of the latter on 
radiobackground, while 
monopolonia and relic particles are concentrated in the Galactic halo, 
and the photon flux dominates. Another signature of the latter sources is 
anisotropy imposed by asymmetric position of the sun in the Galactic halo. 
In all cases considered so far, including necklaces, photons 
must be present in ultra-high energy radiation observed from topological
defects, and experimental discrimination between 
photon-induced and proton-induced extensive air showers can give a clue 
to the origin of ultra-high energy cosmic rays.   

\end{abstract}

\section{Introduction}

The observation of cosmic ray particles with energies higher than $10^{11}~GeV$
\cite{EHE}  gives a serious challenge to the known mechanisms of acceleration. 
The shock acceleration in various
astrophysical objects typically gives maximal energy of accelerated protons
less than $(1-3)\cdot 10^{10}~GeV$ \cite{NMA} (see however \cite{Bier97}). 
The unipolar induction can 
provide the maximum energy $1\cdot 10^{11}~GeV$ only for extreme values 
of the parameters \cite{BBDGP}. Much attention has recently been given to 
acceleration by ultrarelativistic shocks \cite{Vie},\cite{Wax}. The
particles here can gain a tremendous increase in energy,
equal to $\Gamma^2$, at a single reflection, 
where $\Gamma$ is the Lorentz factor of the shock.
However, it is known (see e.g.  the simulation 
for pulsar relativistic wind in \cite{Hosh}) that particles entering 
the shock region are captured there or at least have a small probability 
to escape. 

Topological defects (TD) (for a review see \cite{Book}) can naturally 
produce particles of ultrahigh energies (UHE) well in excess of those 
observed in cosmic rays (CR). In most cases the problem with topological 
defects is not the maximum energy but the fluxes. However, in some cases  
the predicted fluxes are comparable with observations.

Usually, UHE particles appear at the decays of superheavy (SH) particles
produced by TD. (We shall refer to these SH particles as
$X$-particles).  Examples discussed in the literature include
ejection of X-particles from 
superconducting strings, emission of X-particles from cusps or
intersections of ``ordinary'' strings, and production 
of such particles in monopole-antimonopole annihilations. Metastable 
SH particles 
can also be relics of an earlier epoch, produced by a thermal or some
other mechanism in the early Universe.

A rather exceptional mechanism of UHE particle production is given by
radiation
of accelerated monopoles connected by strings. In  this case a monopole 
can radiate gluons with very large Lorentz factors and with virtualities 
of the order of the monopole acceleration.

A common signature of all extragalactic UHECR is the Greisen-Zatsepin-Kuzmin
(GZK) cutoff \cite{GZK}. It reveals itself as a steepening of the spectrum 
of UHE protons and nuclei due to their interaction with microwave radiation. 
The steepening starts at $E \approx 3\cdot 10^{10}~GeV$. Apart from this 
steepening there is another signature of interaction of extragalactic CR 
with microwave radiation: a  bump in the spectrum preceeding the cutoff.
The bump is a consequence of the proton number conservation in the 
spectrum: protons loose energy and are accumulated before the cutoff.

In this paper we will discuss the signatures of UHECR from TD
distinguishing them from particles produced by astrophysical 
accelerators.

We will confine ourselves here to the case of the conventional primary 
particles, protons and photons, and will not consider the other UHE signal 
carriers discussed in the literature such as neutrinos \cite{bz,bordes},
Lightest Supersymmetric Particles \cite{cfk,BeKa}, relativistic monopoles 
\cite{wk} and vortons \cite{bona}.

Throughout the paper we shall use the following numerical values and 
abbreviations: the dimensionless Hubble constant $h=0.65$, the Cold Dark 
Matter density in terms of critical density 
$\Omega_{CDM}=0.2 h^2$, the size of Dark Matter halo $R_h=100~kpc$, 
UHECR - for Ultra High Energy Cosmic Rays and UHE - for Ultra High Energy,
TD - for Topological Defect, SH - for Superheavy, CDM - for Cold Dark 
Matter, SUSY - for Supersymmetry, LLA - for Leading Logarithmic 
Approximation, AGN - for Active Galactic Nucleus, GC and AC - for 
Galactic Center and Anticenter, respectively. 

\section{Constraints and signatures}
A common characteristic feature of UHE particle production in TD is an 
excess of pions over nucleons \cite{Ahar92,SSB}. As a result in many cases 
the observed UHE gamma-ray flux dominates over proton flux and it makes 
$\gamma/p$ ratio ``a diagnostic tool" \cite{Ahar92} for TD. We shall discuss 
this signature quantitatively for various sources, such as cosmic 
necklaces, monopolonia and superheavy relic particles.

An excess of photons over protons {\em at generation} is present in all 
known mechanisms 
of high energy particle generation: decay of SH particles produced by 
strings and cusps, annihilation of monopoles and radiation of UHE particles 
by accelerated monopoles. The order of magnitude of this excess can be 
estimated 
from the ratio of energy transferred to pions and nucleons in the QCD cascade.
For example, $N/\pi$ ratio for the decay $Z^0 \to hadrons$ is about $5\%$
($N$ includes $p,\bar{p},n, \bar{n}$ and $\pi$ -- charged and neutral pions). 
To estimate $\gamma/N$-ratio at energy of interest, one 
should take into account two photons produced by each $\pi^0$-decay, the
fraction (1/3) of neutral pions relative to all pions and the energy spectrum 
of produced hadrons. It gives 
roughly $\gamma/N \sim 10$ at production. However, in the observed diffuse 
flux at $E> 10^6~GeV$
the proton component~\footnote{Here and below we 
shall refer to the $p+\bar{p}$ diffuse flux from TD as to {\em proton} flux.} 
can dominate because of the strong absorption
of high energy photons on the background photon radiation.
UHE photons are absorbed mainly on radio background \cite{B70,PB} and the 
absorption length
is sensitive to low-frequency cutoff in this background.

The pion dominance of hadron production by TD has two consequences.
(i)A large part of hadron energy is transferred, due to pion decays, to an
electromagnetic cascade. This can be used to derive an upper bound on
the UHE proton flux.  
(ii) TD localized inside the sphere of gamma-ray absorption give a 
direct flux of UHE photons at the Earth, producing thus an observable 
anisotropy.

{\em The e-m cascade upper limit} on UHECR arises due to cascading of 
electrons and photons in the Universe down to the observed energies 
$10~MeV - 100~GeV$. The flux of the cascade photons at these energies
must be lower than the extragalactic flux measured by EGRET \cite{EGRET}.
The cascade photon flux is below the EGRET extragalactic flux at 
$10~MeV - 100~GeV$ if the energy density of the cascade photons is 
$\omega_{cas} \leq 2\cdot 10^{-6}~eV/cm^3$. This result follows from 
Monte Carlo simulation in \cite{ps,Sigl96} and from our own estimates 
based on analytic calculations in \cite{BBDGP,B92}). This limit 
(which will be used in our calculations) is rather rigorous: due to 
uncertainties in infra-red flux and intergalactic magnetic field the 
allowed density of the cascade radiation can be as high as  
$(3-5)\cdot 10^{-6}~eV/cm^3$.   

The energy density of
e-m cascades can be readily calculated as  

\begin{equation}
\omega_{cas}=\frac{1}{2}f_{\pi}m_X
\int_0^{t_0}dt\dot{n}_X(t)(1+z)^{-4} ,
\label{eq:cas}
\end{equation}
where $\dot{n}_X(t)$ is the rate of X-particle production at the epoch $t$ 
(redshift $z$),
$t_0=2.06\cdot 10^{17}h^{-1}~s$ is the age of the 
Universe, $h$ is the dimensionless Hubble constant,
$f_{\pi}$ is the fraction of energy transferred to pions at the decay 
of X-particle, and 1/2 takes into account that half of this energy goes into 
e-m cascade.
One can parametrize the effect of the evolution of TD on 
X-particle production as \cite{BHSS}
$\dot{n}_X(t)=\dot{n}_X(t_0)(t/t_0)^{-m}$. In most cases, e.g. ordinary 
strings, monopolonium,  and 
necklaces, $m=3$. In this case the integral in Eq.(\ref{eq:cas}) reduces to
$\int dz/(1+z)^2$, i.e. the evolutionary effects are absent. In the case 
of superconducting strings, $m$ is model-dependent and it can be $m=4$ or 
larger. Weak cosmological effects are present in this case.

With $\omega_{cas}$ from Eq. (1) and the diffuse proton flux being 
determined by
$f_N \dot{n}_X(t_0)$, where $f_N$ is a fraction of $m_X$ transferred to 
UHE nucleons, one can obtain an upper bound on the
diffuse flux of UHE (for another approach see \cite{ps}).

It is easy to generalize the calculations above to the case when 
X-particles are produced with a large Lorentz factor $\Gamma$.

Another general restriction on TD models is given by the distance
between the topological 
defects, $D$.
There are three distance scales in our problem: the distance between TD,
$D$, the photon absorption length, $R_{\gamma}(E)$, and the proton attenuation
length, $R_p(E) =c(E^{-1}dE/dt)^{-1}$, where $dE/dt$ is the energy loss of 
UHE proton on microwave radiation. We shall analyze the case 
$R_{\gamma}<R_p$, though at very high energies, $E>10^{12}~GeV$, they  
can be comparable.

For TD with $D>R_p(E)$ the diffuse flux at a representative point between 
TD is exponentially suppressed. For a power-law generation spectrum
with exponent
$\gamma_g$ and the distance $D/2$ to the nearest source, the
suppression factor
for rectilinear propagation is
\begin{equation}
  \exp\left( -(\gamma_g-1)D/2R_p(E) \right).
\label{eq:suppr}
\end{equation}
In the case of diffusive propagation, $D/2$ should be replaced by the
propagation
time.
Such TD are disfavored as sources of UHECR, because their spectrum either 
has an exponential cutoff at energy where $R_p(E)<D$ or, in case of accidental
proximity of a source, the flux is anisotropic. The anisotropy can be estimated
as $\delta \sim l/2ct$, where $l$ is the distance to the source and $t$ is the 
propagation time.

In the other extreme case, $D<R_{\gamma}(E)$, a TD located inside the 
photonic sphere $R_{\gamma}$ creates a direct UHE photon flux, 
$F_{\gamma}=Q_{\gamma}/4\pi r^2$ at the point of observation. The 
produced
Extensive Atmospheric Showers (EAS) can be identified as photon-induced 
EAS in the direction of the source.

The {\em proton-induced} showers can dominate in the case
$R_{\gamma}(E)<D<R_p(E)$.
The sources might be absent inside the photonic sphere with radius $R_{\gamma}$
and thus, if magnetic field is strong enough, no source is seen directly.
The proton spectrum exhibits the usual GZK cutoff, due to energy losses of 
the nucleons produced by the sources beyond the distance $R_p$. Since 
at extremely high energies $R_p$ and $R_{\gamma}$ are not much different,
the number of sources inside the GZK sphere of radius $R_p$ is not
large and some anisotropy is expected.

Finally let us turn to the case $D<<R_{\gamma}(E)$. Generically, this is a 
case
of uniformly distributed sources. The proton showers dominate when 
$R_p(E)>R_{\gamma}(E)$; their spectrum has the usual GZK-cutoff. A certain 
fraction
of showers, $\sim R_{\gamma}/R_p$, are the photon-induced ones, and they 
correspond to the direct arrival of the photons. The $\gamma/p$-ratio
depends thus on the calculations of attenuation lengths of photons and 
protons. We shall analyze this case quantitatively.

A special situation arises when the  sources are concentrated in the galactic 
halo. This happens in the case of relic superheavy particles \cite{BKV}. 
Here we point out that the same phenomenon of galactic enhancement occurs
for monopolonium and decaying vortons. 

The energy losses and absorption are negligible for the halo model and 
thus photons strongly dominate.

Another signature of this model was indicated recently in \cite{DuTi}:
because of asymmetric position of the Sun in the Galactic halo, there 
is a considerable anisotropy of UHE photon flux.

\section{Fluxes}

In this section we shall give the formulae for UHE proton and photon fluxes 
from extragalactic space and from the halo (for the case of SH relic particles 
and monopolonia). The basic quantity which determines these fluxes is the 
rate of X-particles production $\dot{n}_X$. In the case of an extragalactic flux
we assume that sources are uniformly distributed in space and 
$\dot{n}_X$ does not depend on time. For the galactic 
model we assume that $\dot{n}_X$ is a function of the distance from Galactic 
Center $R$. 

X-particles can be produced by TD at rest or with a Lorentz factor $\Gamma$.
The decay of an X-particle results in a parton cascade. The energy spectrum 
of hadrons outside the confinement radius is described by fragmentation 
functions $W_N(x,m_X)$ for nucleons and $W_{\pi}(x,m_X)$ for pions,
where $x=2E/m_X$ and $E$ is the energy of a proton or a photon. 
For fragmentation functions we use the 
supersymmetric generalization \cite{BK98} of the LLA limiting spectrum  
of QCD cascade \cite{DHMT}  normalized 
by the fraction of energy transferred to the nucleons $f_N$, and pions 
$f_{\pi}$, respectively:
\begin{equation}
\int_0^1 dx x W_i(x,m_X)= 2f_i
\end{equation}
with $i=N,\;\; \pi$. In all calculations below we shall use 
$f_{\pi} \approx 0.5$, as suggested by calculations \cite{BeKa}, which show 
that about half of energy of SUSY-QCD cascade is taken away by neutralinos 
and high-energy leptons. For the ratio $f_N/f_{\pi}$ we shall use $0.05$ 
inspired by data on $Z^0$- decay.  
The SUSY-QCD fragmentation function considerably differs from that of 
ordinary QCD \cite{BK98}: the maximum of the Gaussian peak is shifted towards  
smaller 
$x$, and the peak is much higher, while at larger $x$ SUSY-QCD spectrum is 
below the ordinary QCD spectrum. Since $\omega_{cas}$ is determined by 
large $x$, {\em there are more UHE particles at fixed $\omega_{cas}$ 
in case of 
SUSY-QCD fragmentation function, as compared with ordinary QCD spectrum.}

As it is discussed in \cite{DHMT,BK98,BiSa} the normalization of analytic 
solutions  by conservation of total energy,
results in large uncertainty in the normalization constant. On the other 
hand, the ordinary QCD spectrum (used e.g. in \cite{BiSa}) has the {\em 
the shape } of the spectrum much different from SUSY-QCD spectrum. One 
needs SUSY-QCD Monte Carlo simulation for properly normalized SUSY-QCD 
spectrum.

The spectra of extragalactic UHE protons and photons can be calculated as

\begin{equation}
I_p^{extr}(E)=\frac{\dot{n}_X^{extr}c}{2\pi m_X}
\int_{0}^{\infty}dt W_N(x_g,m_X)
\frac{dE_g(E,t)}{dE},
\label{eq:p}
\end{equation}
\begin{equation}
I_{\gamma}^{extr}(E)=R_{\gamma}(E)\frac{\dot{n}_X^{extr}}{\pi m_X}
\int_{2E/m_X}^{1}\frac{dx}{x}W_{\pi^0}(x,m_X),
\label{eq:gamma}
\end{equation}
\begin{equation}
\frac{dE_g(E,z_g)}{dE}=(1+z_g)exp \left(\int_0^{z_g}\frac{dz}{H_0}(1+z)^{1/
2}(db(E,0)/dE)_{E=E_g(z)} \right),
\label{eq:dif}
\end{equation}
where $x_g=2E_g/m_X$ and $E_g(E,t)$ is the energy of generation at time $t$
for a proton with energy $E$ now. The energy $E_g$ is determined by proton
energy losses 
on microwave radiation $dE/dt=b(E,z)$ and due to redshift.
The expression $dE_g/dE$, given by Eq.(\ref{eq:dif}), is taken from 
\cite{BG88}.
 
The photon absorption length, $R_{\gamma}(E)$, at 
very high energies is determined mostly by pair production on radio 
background. For our calculations we use the absorption lengths from 
\cite{PB}. The photon absorption lengths from \cite{B70} and 
\cite{PB} are plotted in Fig.1. 

As one can see from Eq.(\ref{eq:p}), the proton flux is 
calculated with recoil protons taken into account, while for photons we 
neglect multiplication due to cascade on background photons.
The reason is that electron-positron pairs produced as a result of 
absorption of primary photons on radio-photons, are losing their energies
on radiobackground practically continuously and  most of their energy is lost 
before a collision with a microwave photon - the process responsible 
for the cascade development.

The ratios $\gamma/p=I_{\gamma}(E)/I_p(E)$ as determined by Eq's(\ref{eq:p},
\ref{eq:gamma}) are plotted in Fig.2 as function of energy for 
two extreme cases of absorption length from \cite{PB}. One can see that 
even for the exceptional case of low $\gamma/p$ ratio considered here 
($D \ll R_{\gamma}$), it becomes 
appreciable, $0.5 \leq \gamma/p \leq 2.5$, at energy $3\cdot 10^{11}~GeV$
and increases further with energy.

Let us consider now the halo model.
In cases of SH relic particles \cite{BKV} and monopolonia, their density 
is everywhere proportional to the density of Cold Dark Matter (CDM). Thus,
the ratio of the rates of X-particle production in the halo,  
$\dot{n}_X^h(r_{\odot})$, and in extragalactic space, $\dot{n}_X^{extr}$, 
is given by
\begin{equation}
\frac{\dot{n}_X^h(r_{\odot})}{\dot{n}_X^{extr}}= \frac{\rho_{CDM}^h(r_{\odot})}
{\Omega_{CDM}\rho_{cr}},
\label{eq:halo}
\end{equation}
where $r_{\odot}$ is the distance between the Sun and the Galactic Center, 
$\Omega_{CDM}$ is the CDM density in the extragalactic space in 
units of critical density $\rho_{cr}$ and 
$\rho_{CDM}^h(r_{\odot}) \approx 0.3~GeV/cm^3$ is the local CDM density in 
the halo.

The fluxes of UHE protons ($i=p$) and photons ($i=\gamma$) from the halo 
can be calculated as
\begin{equation}
I_i^h(E)= \frac{W_i(E)}{4\pi}\int_0^{\pi}\cos\theta d\cos\theta
\int_0^{r_{max}}dr\dot{n}_X^h(R),
\label{eq:fh}
\end{equation}

where $\theta$ is an angle relative to the direction of GC and 
$$
r_{max}(\theta)=r_{\odot}\cos\theta+\sqrt{R_h^2-r_{\odot}^2\sin^2\theta}.
$$
The rate of particle production, $\dot{n}_X^h(R)$, is parametrize here as 
density of DM \cite{Prim},
\begin{equation}
\dot{n}_X^h(R)=\frac{\dot{n}_0^h}
{(R/r_0)^{\gamma}[1+(R/r_0)^{\alpha}]^{(\beta-\gamma)/\alpha}},  
\label{eq:dd}
\end{equation}
where $\dot{n}_0^h$ is the normalizing rate, $R$ is the distance from the
Galactic Center, 
$\alpha,\beta,\gamma=(2,2,0)$ correspond to isothermal 
profile, $\alpha, \beta, \gamma = (2,3,0.2)$ gives, according to 
\cite{Prim}, the best fit to observational data and 
$\alpha, \beta, \gamma=(1,3,1)$ describes the numerical simulations of 
Ref.(\cite{NFW}).

The functions $W_p(E)$ and $W_{\gamma}(E)$ in Eq.(\ref{eq:fh}) can be given 
in terms of fragmentation functions, $W_N(x,m_X)$
and $W_{\pi^0}(x,m_X)$, as 
\begin{eqnarray}
W_p(E)=\frac{1}{m_X}W_N(x,m_X),   \nonumber\\
W_{\gamma}(E)=\frac{2}{m_X}\int_{2E/m_X}^1 \frac{dx}{x}W_{\pi^0}(x,m_X).
\label{eq:frag}
\end{eqnarray}

Relic particles and monopolonia fill also extragalactic space with 
space density proportional to $\Omega_{CDM}\rho_{cr}$. The decay rate 
$\dot{n}_X^{extr}$ is given by Eq.(\ref{eq:halo}) and the fluxes by 
Eq's(\ref{eq:p}) and (\ref{eq:gamma}).

\section{Sources}

The following Topological Defects have been discussed
as potential sources of UHE particles.
\begin{itemize}
\item[(i)]
Superconducting strings \cite{HSW},
\item[(ii)]
Ordinary strings \cite{BR}, including the cusp radiation \cite{Bran}
\item[(iii)] 
Networks of monopoles connected by strings \cite{VV87,BMV},
\item[(iv)] 
Necklaces \cite{BV}: hybrid topological defects, where each 
monopole is attached to two strings,
\item[(v)] 
Magnetic monopoles, or more precisely bound monopole-antimonopole
pairs (monopolonium \cite{Hill,BS}).
\item[(vi)]
Vortons \cite{dash}: small loops of superconducting string stabilized
by their angular momentum.
\end{itemize}

Finally we include in this list
SH quasistable relic particles produced in the early Universe 
\cite{KR,BKV,KuTk,Ellis,Fram,BiSa}.
Here we shall apply the criteria discussed in the previous section to
each of these sources.

(i) {\em Superconducting strings}\\

Superconducting strings produce SH particles when the electric current
 in the strings reaches the critical value, $i = i_c$. In some scenarios,
e.g. \cite{OTW}  where the current is induced by primordial magnetic field,
the critical current produces strong magnetic field, in which all high 
energy particles degrade catastrophically in energy \cite{BeRu}. 
However, for {\em ac} currents there are portions of the string with large 
electric charge and small current. High energy particles can escape from
there.

Large {\it ac} currents can be induced in string loops as they oscillate in
 galactic or extragalactic magnetic fields. Even if the string current
 is typically well below critical, super-critical currents can be
 reached in the vicinity of cusps, where the string shrinks by a large
 factor and density of charge carriers is greatly enhanced. In this
 case, SH particles are emitted with large Lorentz factors.

Loops can also acquire {\it dc} currents at the time of formation, when they
are chopped off the infinite strings. As the loops lose their energy
by gravitational radiation, they shrink, the {\it dc} currents grow, and
eventually become overcritical. There could be a variety of
astrophysical mechanisms for
excitation of the electric current in superconducting strings, but for 
all mechanisms considered so far the flux of 
UHE particles is smaller than the observed flux \cite{BeVi}.  However,
the number of possibilities to be explored here is very large, and
more work is needed to reach a definitive conclusion.\\

(ii) {\em Ordinary strings}\\

There are several mechanisms by which ordinary strings can produce UHE 
particles.

For a special choice of initial conditions, an ordinary  loop can collapse to a
double line, releasing its total energy in the form of X-particles\cite{BR}. 
However, as noted in \cite{BR}, the probability of this mode of collapse is
extremely small, and its contribution to the overall flux of UHE
particles is negligible.

String loops can also 
produce X-particles when they self-intersect (e.g. \cite{Shell}).
Each intersection, however, gives only a few
particles, and the corresponding flux is very small \cite{GK}. 

Superheavy particles with large Lorentz factors can be produced in 
the annihilation of cusps, when the two cusp segments overlap \cite{Bran}.  
The energy released in a single cusp event can be quite large, but
again, the resulting flux of UHE particles is too small to account for
the observations \cite {Bhat89,GK}.

One effect 
which was not considered in \cite{GK} and \cite{Bran} could increase
the production of  
UHE particles \cite{HR84,BR}.
As a non-intersecting closed loop
oscillates and radiates away its energy, the loop configuration is
gradually changing. After the loop has lost a substantial part of its
energy, it is likely to self-intersect and fragment into several
smaller loops. These daughter loops will go through the same cycle,
and the process will continue until the size of the fragments becomes
comparable to the string thickness, at which point the fragment loops
disintegrate into relativistic particles. UHE radiation is also
emitted from cusps on the daughter loops.

The process of loop
fragmentation is not well understood.  We do not know, for example,
the number and size distribution of the fragments.  The
only numerical simulation of the fragmentation process that we are
aware of, \cite{SchP}, used initial loops of arbitrary shape, and it is not 
at all
clear that the results are relevant for loops produced in a realistic
evolving network.  To address this problem, we used a simple analytic
model (see Appendix A) which assumes that, in each round of fragmentation, 
a loop looses a
fixed fraction $(1-f_1)$ of its energy to gravitational waves and breaks 
into a fixed number $(N_1+1)$ of daughter loops of roughly equal size.
The daughter loops move with Lorentz factors $\Gamma_1$ in the rest
frame of the parent loop.  This model, which is similar to that
introduced in Ref. \cite{HR84}, is analyzed in Appendix A, with
the conclusion that, for reasonable values of the parameters $f_1$,
$N_1$ and $\Gamma_1$, the UHECR flux from fragmenting loops is still
too small.  While these results are suggestive, a detailed numerical
simulation of loop fragmentation with gravitational back-reaction will
be needed to reach a more definitive conclusion. However, there is one 
general argument against this model of UHECR production. In the chain of 
fragmentation the Lorentz factor of the daughter loops increases 
multiplicatively, e.g. in our model as $\Gamma_1^n$, and at the moment of 
collapse, reaches a large value. The large maximum energy of X-particle 
results, for the observed UHECR flux, in an unacceptably high cascade
gamma radiation.
This fact, observed in all calculations (see for example \cite{ps}) is easy 
to understand. The total energy of decaying X-particle is released 
mostly in the 
particles of highest energies. When one fixes the flux of UHECR at a 
given energy, e.g. $E= 1.0\cdot 10^{20}~eV$, and increases $E_{max}$, 
the energy transferred to e-m cascades increases. This argument works 
especially strongly in case of cusps, where the Lorentz factor can reach 
tremendous values, e.g. $\Gamma \sim 10^{15} - 10^{17}$ in \cite{Bran}.

It has been recently argued \cite{vincent} that long
strings lose most of
their energy not by production of closed loops, as it is generally
believed, but by direct emission of heavy X-particles.
If correct, this claim will change dramatically 
the standard picture of string evolution. It has been also
suggested that the decay products of particles produced in this
way can explain the observed flux of UHE cosmic rays \cite{vincent,ViHiSa}. 
However, we are not convinced that the conclusions of Ref. \cite{vincent} are 
justified. In fact, we believe that numerical simulations described in
\cite{vincent} allow an alternative interpretation.  
The initial string separation in these simulations is comparable to
the string thickness.  As a result,
string intersections and reconnections can generate
a large amount of energy in
the form of short-wavelength perturbations on the strings.  These
perturbations may then be released in the form of particles from
portions of the string that undergo contraction in the course of the
following evolution.

Even if the conclusions of \cite{vincent} were correct, the
particle production mechanism suggested in that paper cannot explain
the observed flux of UHE particles. If particles are emitted directly
from long strings, then the distance between UHE particle sources is
of the order of the Hubble distance, $D \sim t_0 \gg R_p$. According to
the discussion
in the preceeding section, the flux of UHE particles in this case is
exponentially suppressed, or, in the case of accidental proximity of a
string to the observer, the flux is strongly anisotropic. A fine-tuning 
in the position of the observer is needed to reconcile both 
requirements, because long strings are separated by 
a Hubble distance.\\

(iii){\em Network of  monopoles connected by strings}.\\

The sequence of phase transitions
\begin{equation}
G\to H\times U(1)\to H\times Z_N
\label{eq:symm}
\end{equation}
 results in the formation of monopole-string networks in which each monopole 
is attached to N strings. Most of the monopoles and most of the strings belong 
to one infinite network. The evolution of networks is expected to be 
scale-invariant with a characteristic scale
\begin{equation}
d=\kappa t,
\label{eq:length}
\end{equation}
where $\kappa=const$. The scale $d$ gives the average distance between 
monopoles and the typical length of string segments. Each string attached 
to a monopole pulls it with a force equal to the string tension, $\mu \sim 
\eta_s^2$, where $\eta_s$ is the symmetry breaking vev of strings. The
monopoles are 
accelerated and radiate gauge quanta at the rate
\begin{equation}
\frac{dE}{dt} \sim \frac{h^2}{6\pi}a^2\sim \frac{\mu^2}{g^2m^2},
\label{eq:e-rate}
\end{equation}
where $h\sim 2\pi/g$ and $m$ are the monopole charge and mass,
respectively, {g} is the gauge coupling 
and $a \sim \mu/m$ is the monopole acceleration. The GUT monopole has magnetic 
and chromomagnetic charges, $ h_m \sim 2\pi/e$ and $h_s \sim 2\pi/3g_s$, 
respectively, where $e$ is the e-m coupling constant and $g_s$ is the
color coupling 
constant. From Eq.(\ref{eq:e-rate}) it follows that the energy losses
are dominated 
by e-m radiation. Then a simple energy balance analysis gives the value of 
$\kappa$ in Eq.(\ref{eq:length}) \cite{VV87},
\begin{equation}
\kappa \sim \mu/e^2m^2.
\label{eq:kappa}
\end{equation}
The energy of gauge quanta (practically photons and gluons) radiated
by monopoles
can be estimated assuming a rough energy equipartition between the monopole 
and string subsystems. Then the monopoles have a typical energy 
$E \sim \mu d$ and Lorentz factor $\Gamma_M \sim \mu d/m$. If the
mass of gauge quanta (or the virtuality $Q^2$ in the case of gluon) is
smaller than the monopole 
acceleration $a$, the typical energy of gauge quanta is 
$\epsilon \sim \Gamma_M a$ \cite{BMV}; otherwise the production rate of 
massive 
gauge quanta is exponentially suppressed. Gluon production is also
suppressed unless $a > \Lambda_{QCD}$, that is, 
\begin{equation}
\mu/m> 1~GeV.
\label{glucond}
\end{equation}
Thus we have for both photons
and gluons with $Q^2 \leq a^2$
\begin{equation}
\epsilon \sim (\mu/em^2)^2\mu t.
\label{eq:gluon}
\end{equation}
The production rate (per unit volume) of these particles is
\begin{equation}
\dot{n}_i \sim n_M(t)\epsilon^{-1}(dE/dt)_i,
\label{eq:rate}
\end{equation}
where subscript $i$ runs through $\gamma$ for photon and $g$ for gluon.
In particular for gluons we obtain
\begin{equation}
\dot{n}_g \sim \frac{1}{4}\frac{e^8}{g_s^2}\left( \frac{m^2}{\mu t}\right)^4.
\label{eq:g-rate}
\end{equation}          
Let us assume that each gluon with energy $\epsilon$ fragments to hadrons
with a power-law energy spectrum $KE^{-q}$. If we take $q=1.5$ the
cascade limit
will be somewhat weaker than in the case of a more realistic QCD fragmentation
function.
Let us proceed with this favourable case. 
Using the normalized fragmentation function
\begin{equation}
N_p(\epsilon,E)dE= \frac{f_N}{2}\sqrt{\epsilon}E^{-1.5}dE
\end{equation}
and the rate of gluon production $\dot{n}_g$ given by Eq.(\ref{eq:g-rate}),
one obtains for the diffuse proton flux
\begin{equation}
I_p(E)dE=\frac{1}{32\pi}f_N \frac{e^8}{g_s^2}\frac {R_p(E)}{t_0^4}
\left( \frac{m^2}{\mu} \right)^4 \left( \frac{E}{\epsilon} \right)^{-1.5} 
\frac{dE}{\epsilon}.
\end{equation}
The cascade energy density is determined by the e-m radiation of monopoles and 
can be estimated as
\begin{equation}
\omega_{cas}= 4\pi e^4\frac{m^4}{\mu}\frac{1}{t_0^2}.
\end{equation}
Requiring that
$\omega_{cas} \leq \omega_{obs}$, the proton flux at
energy $E \sim 1\cdot 10^{11}~GeV$ is bounded by
\begin{equation}
E^3 I_p(E) <3.7 \cdot 10^{14}\mu_6^{-1}~m^{-2}s^{-1}sr^{-1}eV^2,
\label{eq:nf}
\end{equation}
where  $\mu_6=\mu/10^6~GeV^2$.
The symmetry breaking scale of strings is unlikely to be below the
electroweak scale, $\mu >10^4~GeV^2$, and the flux (\ref{eq:nf}) is
considerably lower than  that observed.\\*[1mm]

(iv){\em Necklaces}.\\

Necklaces are hybrid TD corresponding to the case $N=2$ in 
Eq.(\ref{eq:symm}), i.e. to the case when each monopole is attached to two
strings.  This system resembles ``ordinary'' cosmic strings,
except the strings look like necklaces with monopoles playing the role
of beads. The evolution of necklaces depends strongly on the parameter
\begin{equation}
r=m/\mu d,
\end{equation}
where $d$ is the average separation between monopoles and antimonopoles
along  the strings.
As it is argued in Ref. \cite{BV}, necklaces might evolve to  
configurations with $r\gg 1$, though numerical simulations are needed to 
confirm this conclusion.  
Monopoles and antimonopoles trapped in the necklaces
inevitably  annihilate in the end, producing first the heavy  Higgs and 
gauge bosons ($X$-particles) and then hadrons.
The rate of $X$-particle production is easy to estimate as \cite{BV} 
\begin{equation}
\dot{n}_X \sim \frac{r^2\mu}{t^3m_X}.
\label{eq:xrate}
\end{equation}

Using Eq's(\ref{eq:p}) and (\ref{eq:gamma}) one can calculate the fluxes of 
UHE protons and gammas taking into account the restriction due to cascade 
radiation
\begin{equation}
\omega_{cas}=
\frac{1}{2}f_{\pi}r^2\mu \int_0 ^{t_0}\frac{dt}{t^3}
\frac{1}{(1+z)^4}=\frac{3}{4}f_{\pi}r^2\frac{\mu}{t_0^2}.
\label{eq:n-cas}
\end{equation}

The separation between necklaces is given by \cite{BV} 
$D \sim r^{-1/2}t_0$ for large $r$. Since $r^2\mu$ is limited by cascade 
radiation, Eq.(\ref{eq:n-cas}), one can obtain a lower limit on the 
separation $D$ between necklaces as
\begin{equation}
D \sim \left( \frac{3f_{\pi}\mu}{4t_0^2\omega_{cas}}\right)^{1/4}t_0
>10(\mu/10^6~GeV^2)^{1/4}~kpc,
\label{eq:xi}
\end{equation}
where we used $f_{\pi} \approx 0.5$. Another (weaker) constraint on the 
parameters of the 
model follows from the condition $d \geq \delta_s$, where 
$\delta_s \sim 1/(e \eta_s)$ is the string 
width and $\eta_s$ is the string symmetry breaking scale.
This condition gives $r_{max} \sim  \eta_m/\eta_s$, where $\eta_m$ is 
the monopole symmetry breaking scale. For $\eta_m \sim 10^{16}~GeV$ and 
$\eta_s \sim 10^3~GeV$, one obtains $r_{max} \sim 10^{13}$, which corresponds 
to $D_{min} \sim 1~kpc$.

Thus, necklaces can give a realistic example of the case $D <<R_{\gamma}$, 
when the Universe is uniformly filled by the sources. The 
proton-induced EAS from necklaces strongly dominate over those induced by 
photons at all 
energies except $E> 3\cdot 10^{11}~GeV$ (see Fig.3), where photon-induced 
showers can comprise an appreciable fraction of the total rate.

The spectra of protons and photons from necklaces are shown in Fig.3.
The calculations were performed with the help of Eq.(\ref{eq:p}) and 
Eq.(\ref{eq:gamma}) using the SUSY-QCD fragmentation functions. The dashed,
dotted and solid lines correspond to the masses of X-particles 
$10^{14}~GeV,\;\; 10^{15}~GeV$ and $10^{16}~GeV$, respectively. The values 
of $r^2\mu$ used to fit these curves to the data are 
$7.1\cdot 10^{27}~GeV^2,\;\;6.0\cdot 10^{27}~GeV^2$ and 
$6.3\cdot 10^{27}~GeV^2$, respectively. They correspond to the cascade 
density $\omega_{cas}$ equal to 
$1.5\cdot 10^{-6}~eV/cm^3,\;\;
1.2\cdot 10^{-6}~eV/cm^3$ and $1.3 \cdot 10^{-6}~eV/cm^3$, respectively, all 
less than the allowed cascade energy density for which we adopt the 
conservative value $\omega_{cas}=2\cdot 10^{-6}~eV/cm^3$.  

The absorption of gamma-radiation on the radio background is taken 
according with the 
upper and lower limits of \cite{PB}, shown in Fig.2. 
The agreement with observational data at the highest energies is improved 
when the photon flux is added to the proton flux.

Two following effects are expected to improve the agreement with the data at 
$E> 1\cdot 10^{11}~GeV$.

The fluctuations in energy losses increase the fluxes. This effect is 
significant when the fraction of energy lost in one collision becomes 
appreciable, i.e. at $E >1\cdot 10^{11}~GeV$. This effect can be accounted 
for in Monte Carlo simulations.

The local enhancement of the density of sources, e.g. within Local 
Supercluster (LS), makes the spectrum at $E > 3\cdot 10^{10}~GeV$ flatter
(see calculations in \cite{BeGr}). The accumulation of necklaces within 
LS is expected for large $r$. Indeed, for sufficiently large $r$, the typical velocity of 
necklaces $v = c/\sqrt{r}$ is less than the escape 
velocity from the 
Local Supercluster and thus necklaces are confined within LS. For $r > 10^{7}$ 
necklaces can also cluster on the galactic scale.

For energy lower than $1\cdot 10^{10}~GeV$ we assume the presence of 
another component with a cutoff at $E\sim 1\cdot 10^{10}~GeV$. It can be 
generated, for example, by jets from AGN \cite{Bier}, which naturally 
have a cutoff at this energy. \\

(v) {\em Monopolonium and SH relic particles.}\\

These two sources exhibit the same clustering property: they act as
non-dissipative matter which clusters in the Universe in the same way as 
Cold Dark Matter.  As a result the density of these particles in the  Galactic
halo is enhanced according to Eq.(\ref{eq:halo}).\footnote{Clustering
of free monopoles in the galactic halo has been briefly 
discussed in Ref. \cite{DuTi}.  The authors argue that the monopole
density $n_M$ is proportional to the dark matter density and that the
$X$-particle production rate due to monopole-antimonopole annihilation
is ${\dot n}_X \propto n_M^2$.  There are, however, some problems with
this picture.  The motion of free monopoles is strongly affected by
the galactic magnetic field, and their density is not likely to follow
that of CDM.  Moreover, the probability for annihilation of free
monopoles in  the galactic halo is extremely small, and the resulting
flux of cosmic rays is negligible for all reasonable values of the
monopole density $n_M$.}  The spectra of UHE 
protons and photons from decays of relic particles in the halo are calculated 
using Eq.(\ref{eq:fh}) for the different distributions  
of X-particles in the halo given by Eq.(\ref{eq:dd}). The results are 
displayed in Fig.4. The mass of X-particle is $m_X= 1\cdot 10^{14}~GeV$. 
The solid, dotted and dashed curves correspond to 
the distribution ({\ref{eq:dd}) with $(\alpha,\beta,\gamma)$ equal to 
$(2,2,0),\;\;(2,3,0.2)$ and $(1,3,1)$, respectively. The shape of the 
spectra is naturally the same for all three curves. Since we normalize the 
spectra by the value $\dot{n}_0^h$  to fit the observational data at 
$E_{\gamma} \sim 10^{10}~GeV$, the fluxes are the same in all three 
cases. The corresponding values of $\dot{n}_0^h$ are 
$6.3\cdot 10^{-42}~cm^{-3}s^{-1},\;\; 1.3\cdot 10^{-41}~cm^{-3}s^{-1}$ and 
$1.6\cdot 10^{-41}~cm^{-3}s^{-1}$, respectively. These values also normalize, 
through relation Eq.(\ref{eq:halo}), the flux of extragalactic 
protons, given by  Eq.(\ref{eq:p}).
The obtained results do not significantly differ from those in \cite{BKV}.
The photon flux is greater than the proton flux by a factor of 6.


Another signature of this model is the anisotropy caused by 
asymmetric position 
of the Sun in the Galactic halo \cite{DuTi}. The anisotropy reveals itself most 
significantly as the large ratio of fluxes in the directions of the Galactic 
Center (GC) and the Galactic Anticenter (GA). We define the anisotropy $A(\theta)$ 
as  this ratio with the fluxes measured within a solid angle limited by angle 
$\theta$ with the line connecting the Sun and the GC. This can be readily calculated 
using  the distribution of the sources in the halo given by Eq.(\ref{eq:dd}).

The anisotropy $A(\theta)$ is plotted in Fig.5 
as a function of $\theta$ for all three density profiles given by  
Eq.(\ref{eq:dd}) with $r_0=5~kpc$ (solid curves) and $r=10~kpc$ (dashed 
curves). The estimated anisotropy is 
relevant for energies $E>1\cdot 10^{10}~GeV$, where the contribution of 
the isotropic lower-energy component is small. 

There is one more signature for this model, given by a direct flux of 
protons from the Virgo cluster, provided that 
they are weakly deflected by magnetic fields in the Local Supercluster (see 
\cite{BKV}). Note that UHE photons from this source are absorbed, and cascade 
radiation does not propagate rectilinearly because of the cascade electron 
deflection in the magnetic field.\\

(v) {\em Vortons}\\

Vortons are charge and current carrying loops of superconducting
string stabilized by their angular momentum \cite{dash}.  Although
classically stable, vortons decay by gradually losing charge carriers
through quantum tunneling.  Their lifetime, however, can be greater
than the present age of the universe, in which case the escaping
$X$-particles will produce a flux of cosmic rays.  The $X$-particle
mass is set by the energy scale $\eta_X$ of string superconductivity.  

The number density of vortons formed in the early universe is rather
uncertain.  According to the analysis in Ref.\cite{BCDT}, vortons are
overproduced in models with $\eta_X > 10^9 GeV$, so all such models
have to be ruled out.  In that case, vortons cannot contribute to the
flux of UHECR.  However, an alternative analysis \cite{mash} suggests
that the excluded range is $10^9 GeV <\eta_X < 10^{12}GeV$, while for
$\eta_X \gg 10^{12}GeV$ vorton formation is strongly suppressed.  This
allows a window for potentially interesting vorton densities
with\footnote{These numbers assume that strings are formed in a
first-order phase transition and that $\eta_X$ is comparable to the
string symmetry breaking scale $\eta_s$.  For a second-order phase
transition, the forbidden range widens and the allowed window moves
towards higher energies \cite{mash}.}
$\eta_X \sim 10^{12}-10^{13}GeV$.  
Production of UHE particles by decaying vortons was studied in 
Ref.\cite{mas}. As we already mentioned above, vortons cluster in the
Galactic halo and the discussion in the preceeding subsection is
therefore directly applicable to this case as well.

\section{Discussion}

We studied observational constraints on various TD models, as well as
possible signatures of TD as sources of the observed UHE radiation
($E\geq 10^{10}~GeV$). 

The most stringent constraint is due to electromagnetic cascades. It 
depends on the energy spectrum of particles from decays and on 
astrophysical quantities which determine the development of the 
cascade (most notably on the flux of 
intergalactic infra-red/optical radiation and on intergalactic magnetic 
field ). The SUSY-QCD spectrum makes the cascade constraint weaker, because 
this spectrum predicts less higher energy particles and more low energy 
particles as compared with ordinary QCD spectrum. In case of very large 
$m_X$ it means that for a given UHECR flux, less energy is transferred to 
the e-m cascade radiation. There are considerable uncertainties in 
the extragalactic flux of infra-red radiation and extragalactic magnetic 
field. The conservative limit on the energy density of the cascade 
radiation imposed the latest EGRET data is 
$\omega_{cas} \approx 2\cdot 10^{-6}~eV/cm^3$. It could be 
$3 - 5\cdot 10^{-6}~eV/cm^3$ with astrophysical uncertainties mentioned 
above. The further progress in the study of origin of EGRET extragalactic 
flux and in calculation of SUSY-QCD spectrum, in the pessimistic case, 
can exclude such 
TD as e.g. necklaces as the sources of observed UHECR.

Another important constraint arises from the fact that at ultra-high
energies, the proton attenuation length
$R_p(E)$ and the photon absorption length $R_{\gamma}(E)$ are both
small compared to the Hubble radius.  Models in which the typical
distance between defects is $D>>R_p$ are disfavored.  In such models, 
the observed 
spectrum would have an exponential cutoff, unless a source is
accidentally close to the observer.  In the latter case the flux would
be strongly anisotropic.  

Finally, in many cases TD give UHECR fluxes lower than the observed ones. 
We showed here that this is the case for monopole-string networks.
Superconducting and ordinary cosmic strings probably belong to this
category as well, although some loopholes still remain to be closed.


With all these constraints taken into account, it appears that
only necklaces, monopolonium and relic SH particles survive as potential 
UHE sources.  

The most important observational signature of TD as sources of UHE CR
is the presence of photon-induced EAS. For all known mechanisms 
of UHE particle production 
the pions (and thus photons) dominate 
over nucleons. At energies lower than $1\cdot 10^{12}~GeV$,
protons have considerably larger attenuation length than photons and the 
observed
proton flux can be dominant. Nevertheless, even in this case  
photons reach an observer from sources located inside  the sphere 
of radius $R_\gamma (E)$ (assuming that $R_\gamma >D$). 
Unlike protons, photons propagate rectilinearly,
indicating the direction to the sources. 

{\em Necklaces} with a large value of $r=m/\mu d > 10^7$ have a small 
separation $D <R_\gamma$.
They are characterized by a small fraction, $R_{\gamma}/R_p$, of
photon-induced  
EAS at energies $10^{10} - 10^{11}~GeV$. This fraction increases with
energy and becomes considerable at the highest energies. For
smaller values of $r \sim 10^4 - 10^6$, when the separation is larger
than $R_{\gamma}$ but still smaller than $R_p$, most of UHE particles
are expected to be protons
(with a chance of incidental proximity of a source, seen as a
direct gamma-ray source). Thus, in all cases necklaces are characterized by
an excess of proton-induced showers.  However, some fraction of photon-induced 
showers is always present, and it can be large at the highest 
energies.

{\em Monopolonium}, decaying vortons {\em and SH relic particles} are
characterized  by an enhanced 
density in the Galactic halo. They give a photon-dominated flux
without a 
GZK cutoff. Due to asymmetric position of the Earth in the Galaxy, this 
flux is anisotropic. The largest flux is expected from the direction of 
the Galactic Center, where the density of sources 
is the largest. Unfortunately, the Galactic Center is not seen by gigantic 
arrays , such as Akeno, Fly's Eye, Haverah Park, and Yakutsk array. However, 
these detectors can observe a minimum in the direction of the Galactic 
anticenter in particles with energies $E>1\cdot 10^{10}~GeV$, as compared 
with the direction perpendicular to the Galactic Plane.

A flux from the Virgo cluster might be another signature of this 
model.

The search for photon induced showers is not an easy experimental task.
It is known (see e.g. Ref.\cite{AK}) that in the UHE photon-induced showers 
the 
muon content is very similar to that in proton-induced showers. However,
some  difference in the muon content between these two 
cases is expected and may be used to distinguish between them
observationally.  A detailed analysis would be needed to determine
this difference. 

The Landau-Pomeranchuk-Migdal (LPM) effect \cite{LPM} and the absorption of 
photons in the geomagnetic field are two other 
important phenomena which affect the detection of UHE photons 
\cite{AK,Kasa}; (see \cite{ps} for a recent discussion). 
The LPM effect reduces the cross-sections 
of electromagnetic interactions at very high energies. However, if the
primary photon approaches the Earth in a direction characterized by a large
perpendicular component of the geomagnetic field, the photon is likely
to
decay into electron and positron \cite{AK,Kasa}. Each of them emits 
a synchrotron photon,
and as a result a bunch of photons strikes the Earth atmosphere. The LPM 
effect, which strongly depends on energy, is thus suppressed. If on the other 
hand
a photon moves along the magnetic field, it does not decay, and LPM effect 
makes shower development in the atmosphere very slow. At extremely 
high energies the maximum of the showers can be so close to the Earth surface 
that it becomes "unobservable" \cite{ps}. 

We suggest that for all energies above the GZK cutoff the showers be analyzed
as candidates for being induced by UHE photons, with the probability
of photon splitting 
in the geomagnetic
field determined form the observed direction of propagation, 
and with the LPM effect taken into 
account. The search for photon-induced showers can be especially effective 
in the case of Fly's Eye detector which can measure the longitudinal
development of EAS. The future Auger detector will have, probably, the 
highest potentiality to resolve this problem.
\section*{Acknowledgements}
We thank Michael Kachelriess for help in calculations and Motohiko Nagano for 
information and interesting discussion. We are grateful to Svetlana Grigorieva 
for providing us with the results of her recent new calculations (unpublished) 
of proton energy losses and the values of $db(E)/dE$ used in 
Eq.(\ref{eq:dif}). A.V. is grateful to Anne Davis and Paul
Shellard for a discussion of vortons.  
We also thank Guenter Sigl for a valuable remark and 
Pijushpani Bhattacharjee for a letter of discussion.

The work of A.V. was supported in part by the National Science Foundation. 
P.B. was supported by a INFN fellowship at the University of Chicago.

\appendix 
\section{UHE particle production due to multiple loop fragmentation}

We shall adopt the following simple model of loop fragmentation (It is
somewhat similar to the model introduced in Ref. \cite{HR84}). Each loop 
fragments into ($N_1 + 1$) daughter loops after it radiated away a fraction 
($1 - f_1$) of its energy. The Lorentz factor of the daughters in the 
center-of-mass frame of the parent
loop is $\Gamma_1$. If the initial mass of the loop is $M$, then after $n$ 
rounds of fragmentation, the number of daughters is
\begin{equation}
N_n \sim N_1^n,
\label{A1}
\end{equation}
and their energy, rest mass and Lorentz factor are, respectively,
\begin{equation}
E_n \sim (f_1/N_1)^nM,
\label{A2}
\end{equation}
\begin{equation}
M_n \sim (f_1/N_1\Gamma_1)^nM,
\label{A3}
\end{equation}
\begin{equation}
\Gamma_n \sim \Gamma_1^n.
\label{A4}
\end{equation}
The fragmentation process stops at round $n_*$ when $M_{n_*} \sim \eta$, where 
$\eta$ is the symmetry breaking scale of strings, that is, when the size of 
the fragments becomes comparable to the string thickness,
\begin{equation}
n_* \sim {\ln (M/\eta) \over{\ln (N_1\Gamma_1/f_1)}}.
\label{A5}
\end{equation}

Loops decaying at the present time $t_0$ have masses $M \sim 10^2G\mu^2t_0$ and
\begin{equation}
M/\eta \sim 10^{55}\eta_{16}^3,
\end{equation}
where $\eta_{16} \equiv \eta/10^{16} GeV$. The typical values of
$f_1$, $N_1$ and $\Gamma_1$ are not known.  Numerical simulations of
loop fragmentation in Ref. \cite{SchP} found $N_1 \sim 3-10$
and $\Gamma_1 \sim 1.3$. However, these simulations used loops of
arbitrary shape, and the parameter values for loops in a realistic
network may be quite different.  It seems reasonable to assume that on
average the loop has to lose at least 30\% of its energy to initiate
the next round of fragmentation, that is, $f_1 < 0.7$. 
As we shall see from Eq.(\ref{A7}), the energy output of loops in the form of 
cosmic rays is maximized for the largest possible values
of $N_1$ and $f_1$.
In the estimates below, we shall adopt $N_1 \sim 10$,
$f_1 \sim 0.7$ and $\Gamma_1\sim 1.3$.  The values of $N_1$ and $f_1$
appear somewhat large, and we shall
keep in mind that we have made a rather optimistic choice of the
parameters.  With these values,
\begin{equation}
n_* \approx 44 + \ln\eta_{16}.
\label{A6}
\end{equation}

Each fragmenting loop gives $\sim N_1$ particles of energy $\sim \eta\Gamma_1$ 
in the rest frame of the loop ($\sim 1$ particle per intersection). The 
fraction $F$ of the total energy of the initial loop that ends up in the form 
of UHE particles can be estimated as
\begin{equation}
F \sim {\eta \over{M}}\sum_{n=1}^{n_*} N_n\Gamma_n.
\end{equation}
The dominant contribution to the sum is given by the last term (that is, by 
the last round of fragmentation), and we can write
\begin{equation}
F \sim {\eta \over{M}}(N_1\Gamma_1)^{n_*} \sim f_1^{n_*},
\label{A7}
\end{equation}
where in the last step we have used the definition of $n_*$. Using
Eq. (\ref{A6}) and $f_1 \approx 0.7$, we have for the fraction of energy 
transferred to X-particles
\begin{equation}
F \sim 2\cdot 10^{-7}\eta_{16}^{-0.4}.
\label{A8}
\end{equation}
Most of the particles are emitted at energies
\begin{equation}
E_X \sim \Gamma_1^{n_*}\eta \sim 9 \cdot 10^{20}\eta_{16}^{0.7} GeV.
\label{A9}
\end{equation}

The X-particle injection rate is given by
\begin{equation}
\dot{n}_X \sim F{\mu \over{E_Xt_0^3}} \sim 3\cdot 10^{-55}
\eta_{16}^{0.9} cm^{-3}s^{-1}. 	
\label{A10}
\end{equation}
To maximize the diffuse proton flux, $I_p(E)$, we shall assume a power-law 
fragmentation function $KE_r^{-p}$ with $p = 1.5$ in the system where 
X-particle is at rest. After simple calculations 
using the Lorentz-transformation to the laboratory system, one obtains 
the diffuse flux as  
\begin{equation}
I_p(E)=\frac{(2-p)f_N}{4\pi p}\frac{\dot{n}_X}{E_X}
\left(\frac{E}{E_X}\right)^{-p} R_p(E),
\label{A11}
\end{equation}
where $R_p(E)$ is the proton attenuation length. 
For $E=6.3\cdot 10^{19}~eV$, using Eqs.(\ref{A9},\ref{A10}), we obtain
$E^3I_p(E) \approx 3\cdot 10^{18} \eta_{16}^{1.25}~eV^2m^{-2}s^{-1}sr^{-1}$ 
to be 
compared with the observed value $3\cdot 10^{24}~eV^2m^{-2}s^{-1}sr^{-1}$. 
Thus the calculated flux is too small by a factor of $10^6$. 
This discrepancy is difficult to resolve by stretching the range of the 
parameters $f_1$ and $N_1$. For example, with $f_1 \sim 0.9$ 
(which appears unreasonably large), the proton flux is 
$E^3I_p(E) \approx 3\cdot 10^{23}\eta_{16}^{1.5}~eV/m^2 s sr$, but the 
cascade energy density is too large  due to increase of $E_X$ :
$$
\omega_{cas} \approx (1/2)E_X\dot{n}_Xt_0 
\approx 7.4\cdot 10^{-3} \eta_{16}^{1.9}~eV/cm^3.
$$

We next consider particle production by cusp evaporation in the fragmenting 
loops. Assuming that cusps are periodically repeated and completely 
"evaporated" into particles, the energy rate of particle production (in the 
rest frame of the loop) is \cite{Bran}
\begin{equation}
{\dot{\cal E}}_p \sim \mu({M \over{\eta}})^{-1/3}.
\label{A12}
\end{equation}
Compared to the gravitational radiation power, 
${\dot{\cal E}}_g \sim 10^2G\mu^2$,  this is
\begin{equation}
{\dot{\cal E}}_p/{\dot{\cal E}}_g 
\sim 10^{-2}(G\mu)^{-1}({M \over{\eta}})^{-1/3}.
\label{A13}
\end{equation}
The Lorentz factor at the cusp is
\begin{equation}
\Gamma_c \sim ({M \over{\eta}})^{1/3}.
\end{equation}
The fraction of energy lost by a fragmenting loop in the form of particles is
\begin{equation}
F = 10^{-2}(G\mu)^{-1}(1 - f_1)
\sum_{n=0}^{n_c}N_n{E_n\over{M}}({M _n\over{\eta}})^{-1/3}
\end{equation}
\begin{equation}
= 10^{-2}(G\mu)^{-1}(1 - f_1)({M \over{\eta}})^{-1/3}
\sum_{n=0}^{n_c}(f_1^2N_1\Gamma_1)^{n/3},
\label{A13bis}
\end{equation}
where
\begin{equation}
n_c \sim 32 + 3\ln \eta_{16}
\end{equation}
is the value of n at which ${\dot{\cal E}}_p \sim {\dot{\cal E}}_g$.

For our choice of parameters, $f_1^2N_1\Gamma_1 > 1$ and the dominant 
contribution to (\ref{A13bis}) is given by $n \sim n_c$. Then
\begin{equation}
F \sim (1 - f_1)f_1^{n_c} \sim 3 \cdot 10^{-6}\eta_{16}^{-1},
\end{equation}
which is comparable to Eq. (\ref{A8}). The energy of the emitted particles is
\begin{equation}
E_X \sim \eta\Gamma_1^{n_c}\Gamma_c \sim 4 \cdot 10^{23}\eta_{16}^{-0.2} GeV.
\end{equation}
This energy is too high: the observed flux cannot be obtained for any
reasonable value of $\eta$ without violating the cascade bound.

\newpage

{\bf Figure Captions}
\vskip .5cm
Fig. 1: Gamma-ray absorption length as a function of energy. The solid
line and the dotted line reproduce the upper and lower case 
of ref. \cite{PB}, while the dash-dotted line is taken from 
ref. \cite{B70} and drawn here for comparison.\par

\vskip .5cm

Fig. 2: The $\gamma/p$ ratio for different values of $m_X$ ( indicated
in the figure) as a function of the gamma-ray energy. High ($\gamma$-high) 
and low ($\gamma$-low) photon fluxes correspond to two extreme cases 
of gamma-ray absorption from \cite{PB}. \par
\vskip .5cm

Fig. 3: Proton and gamma-ray fluxes from necklaces. High ($\gamma$-high) 
and low ($\gamma$-low) photon fluxes correspond to two extreme cases 
of gamma-ray absorption from \cite{PB}. The fluxes are given for 
$m_X=1\cdot 10^{14}$ GeV (dashed lines), $m_X=1\cdot 10^{15}$ GeV (dotted 
lines) and
$m_X=1\cdot 10^{16}$ GeV (solid lines). The fluxes are normalized to the 
observed data.\par
\vskip .5cm

Fig.4: Predicted fluxes from relic SH particles ($m_X=1\cdot 10^{14}~GeV$) or 
from monopolonia producing X-particles with the same masses: nucleons from
the halo (curves labelled as ``protons''), gamma-rays from the halo (curves
labelled ``gammas'') and extragalactic protons (as indicated). The solid,
dotted and dashed curves correspond to $(\alpha,\beta,\gamma)=
(2,2,0),~(2,3,0.2)$ and $(1,3,1)$ respectively (see text).\par
\vskip .5cm

Fig. 5: Anisotropy $A(\theta)$ as a function of the angle $\theta$. 
$A(\theta)$ is defined as a ratio of fluxes in the direction of the GC and 
AC. The fluxes are calculated within solid angles limited by angle 
$\theta$ with the line connecting the Sun and GC. Anisotropy is given for 
three density profiles : $(\alpha,\beta,\gamma)=(2,2,0)$ - ``isothermal''
curve, $ (2,3,0.2)$ - ``best fit'' curve, and $(1,3,1)$ - for numerical 
simulation shape.  The solid lines
correspond to $r_0=5$ kpc and the dashed lines to $r_0=10$ kpc.\par

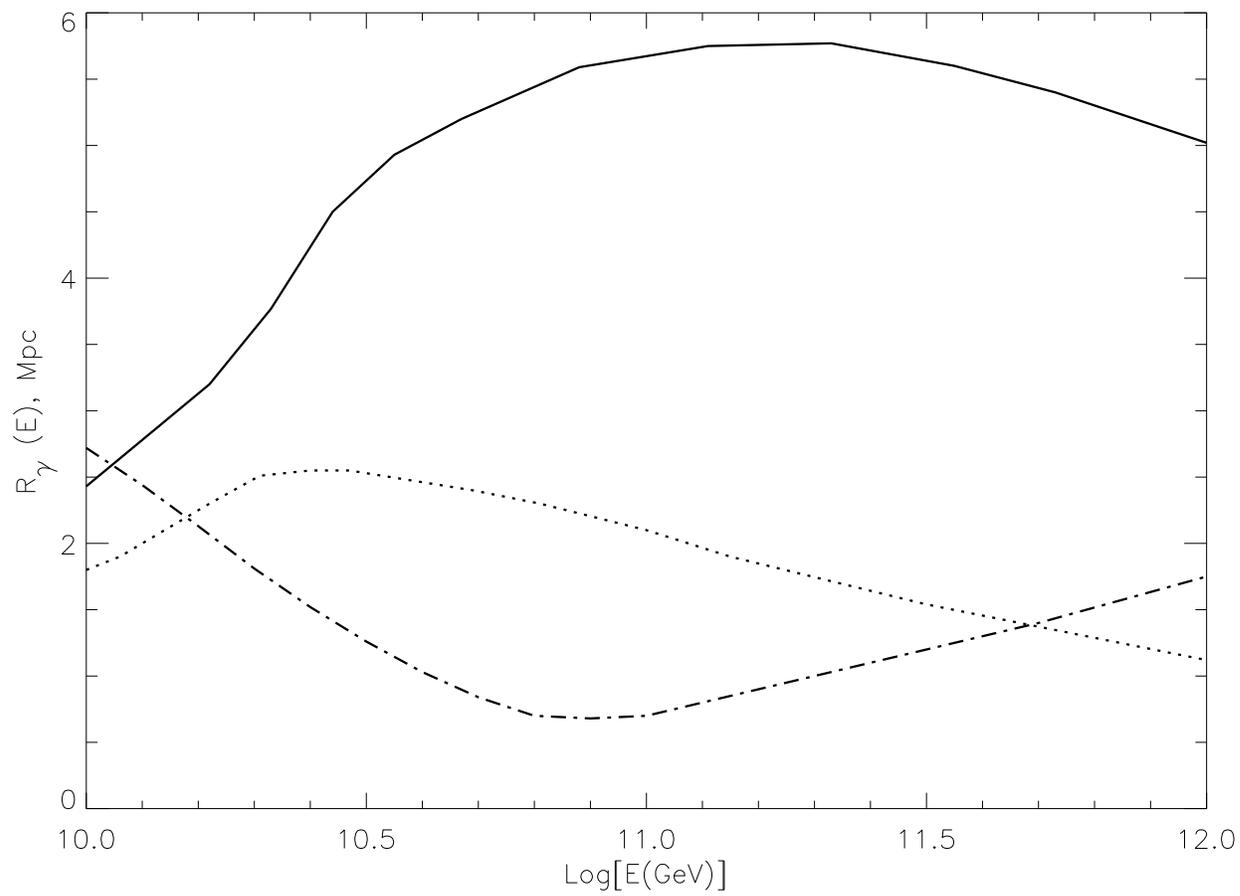
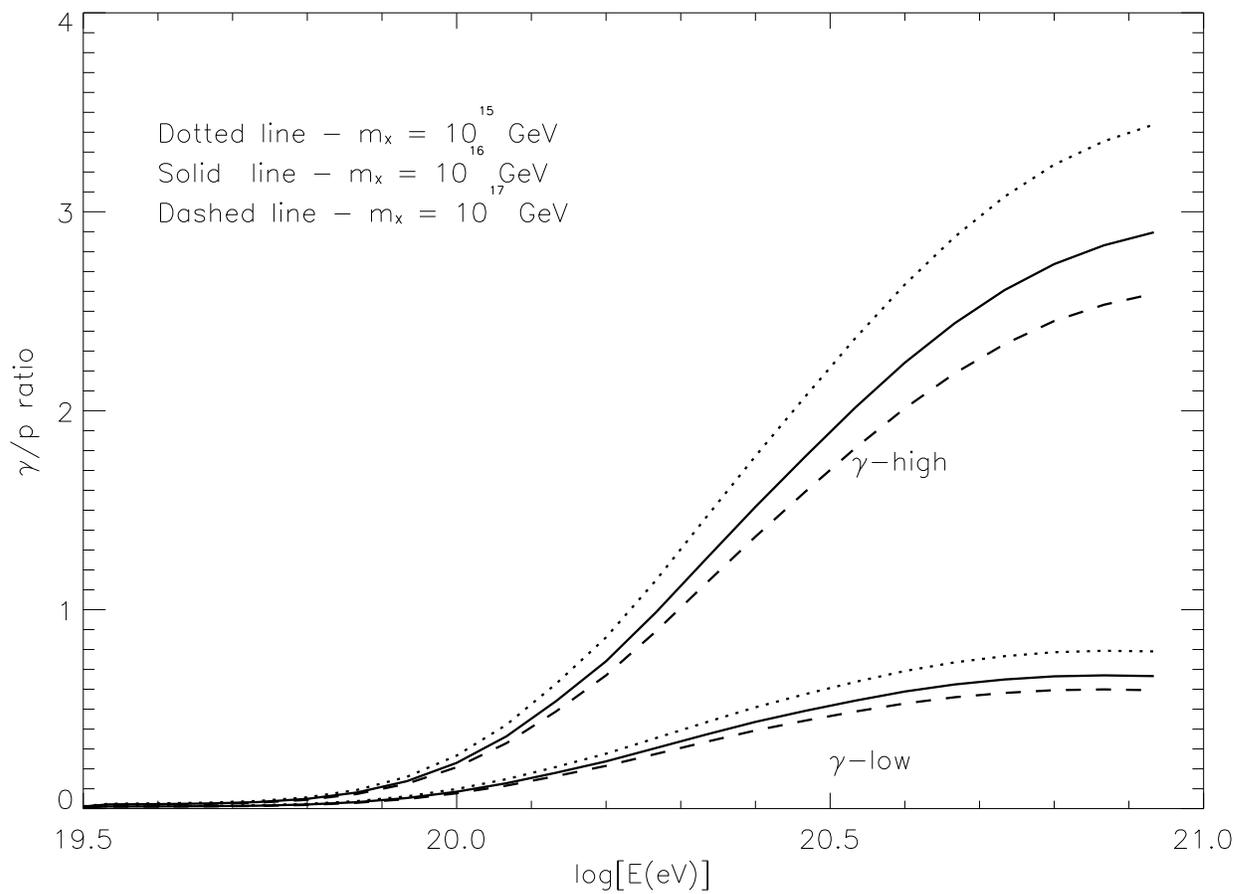
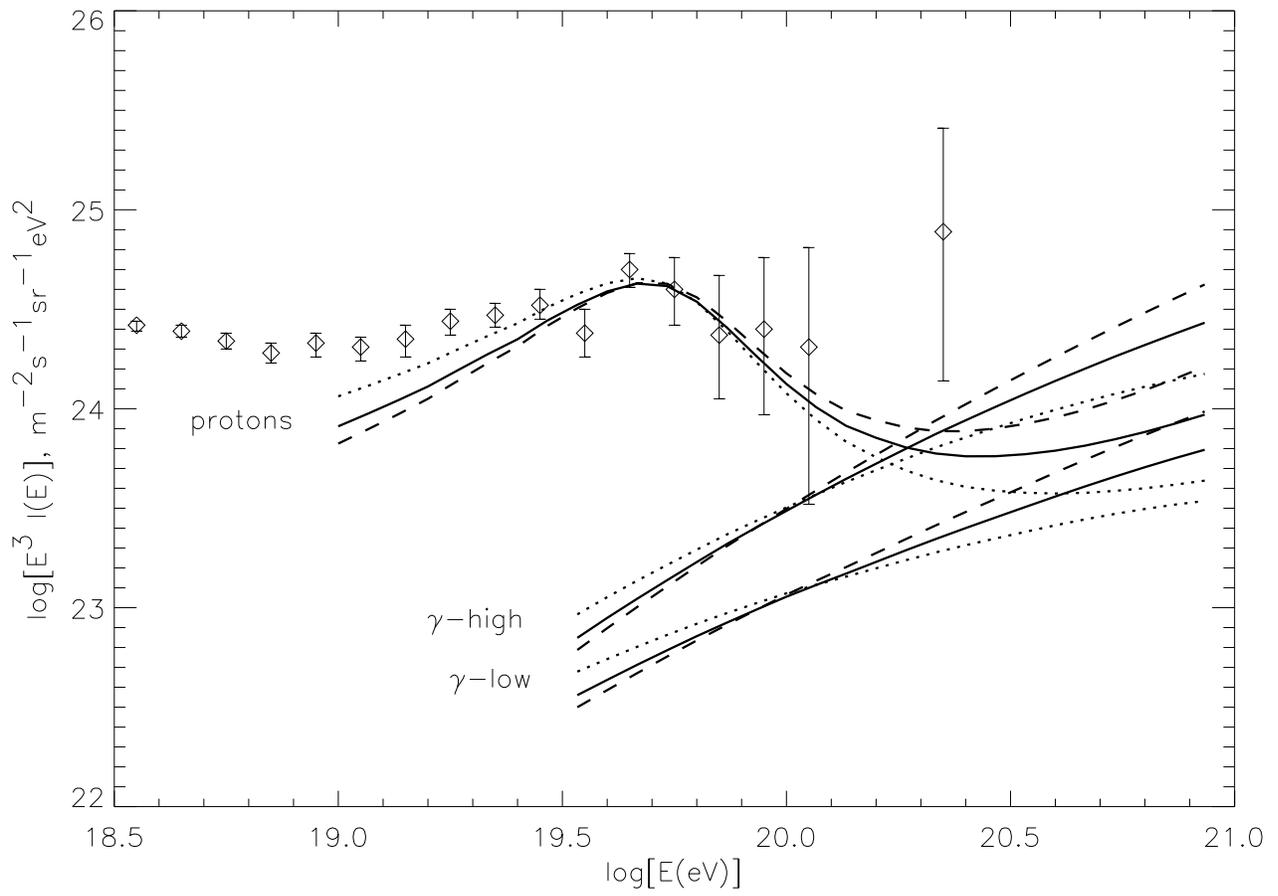
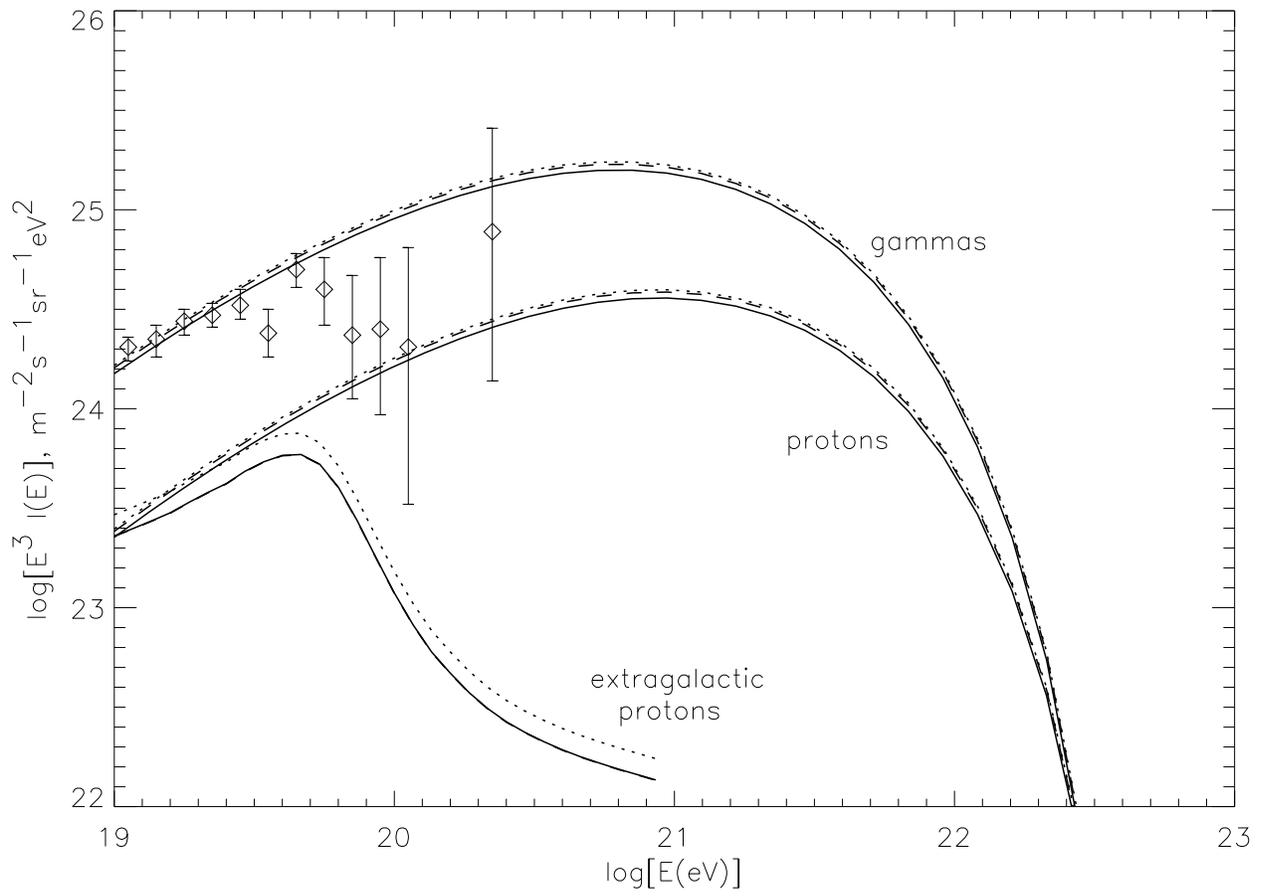
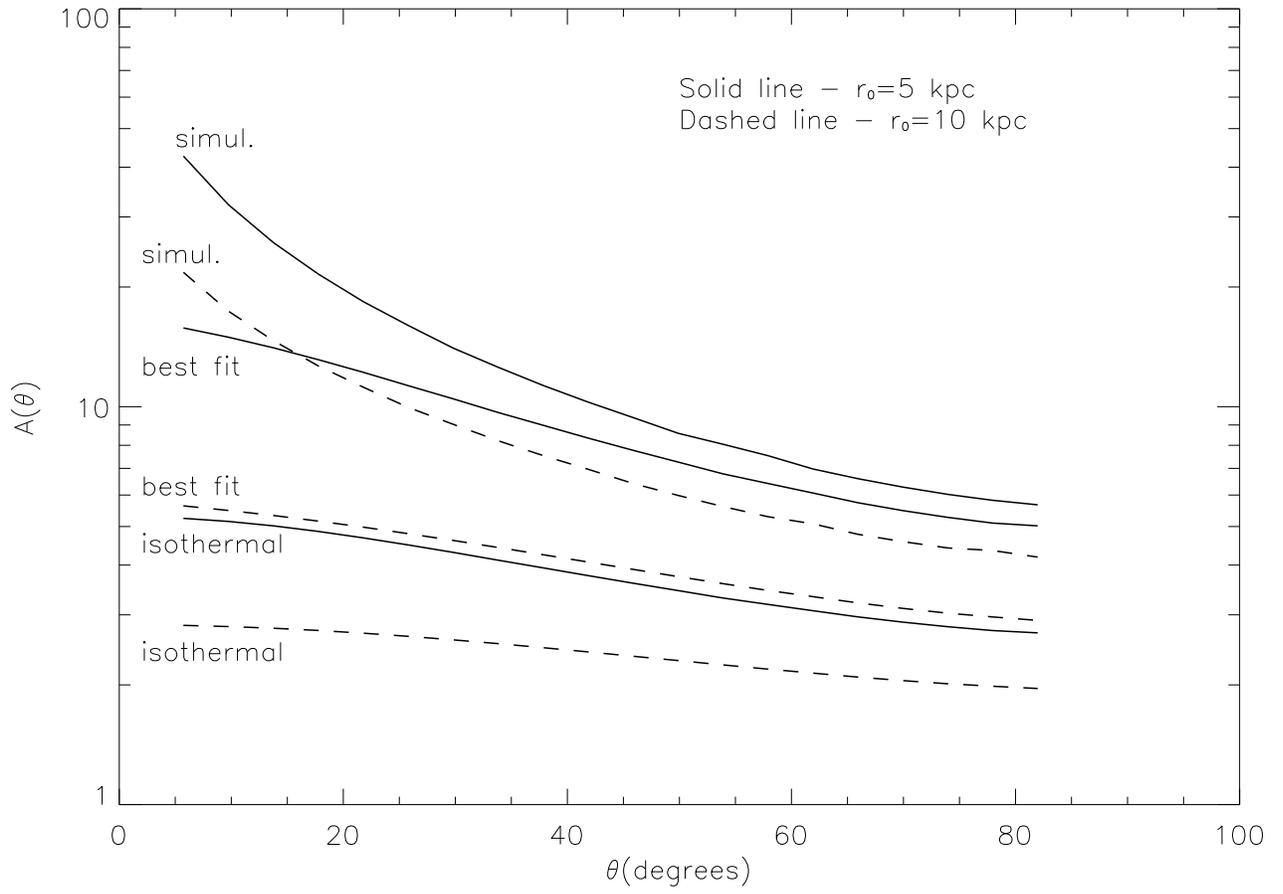
\end{document}